\documentclass[10pt]{article}
\usepackage[utf8]{inputenc}
\usepackage{indentfirst}
\usepackage[english]{babel}
\usepackage[]{amsthm}
\usepackage[]{amssymb}
\usepackage{amsmath}
\usepackage{appendix}
\usepackage{geometry}
\usepackage{graphicx}
\usepackage{hyperref}
\usepackage{subcaption}

\DeclareCaptionSubType{figure}

\usepackage{float}
\usepackage[T1]{fontenc}
\usepackage{authblk}
\usepackage{bm}
\usepackage[shortlabels]{enumitem}
\usepackage{tcolorbox}
\usepackage{tikz}
\usetikzlibrary{matrix}
\makeatletter

\newcommand{\Rmnum}[1]{\expandafter\@slowromancap\romannumeral #1@}
\makeatother

\begin{document}
 
\title{Exploring Implied Certainty Equivalent Rates in Financial Markets: Empirical Analysis and Application to the Electric Vehicle Industry}
\author{Yifan He}
\author{Svetlozar Rachev}

\affil{Department of Mathematics and Statistics, Texas Tech University, Lubbock, TX 79409-1042, USA\authorcr\href{mailto:yifan.he@ttu.edu}{yifan.he@ttu.edu}; \href{mailto:zari.rachev@ttu.edu}{zari.rachev@ttu.edu}}

\renewcommand*{\Affilfont}{\small}
\renewcommand\Authands{ and } 

\date{\today}
\maketitle

\begin{abstract}
In this paper, we mainly study the impact of the implied certainty equivalent rate on investment in financial markets. First, we derived the mathematical expression of the implied certainty equivalent rate by using put-call parity, and then we selected some company stocks and options; we considered the best-performing and worst-performing company stocks and options from the beginning of 2023 to the present for empirical research. By visualizing the relationship between the time to maturity, moneyness, and implied certainty equivalent rate of these options, we have obtained a universal conclusion---a positive implied certainty equivalent rate is more suitable for investment than a negative implied certainty equivalent rate, but for a positive implied certainty equivalent rate, a larger value also means a higher investment risk. Next, we applied these results to the electric vehicle industry, and by comparing several well-known US electric vehicle production companies, we further strengthened our conclusions. Finally, we give a warning concerning risk, that is, investment in the financial market should not focus solely on the implied certainty equivalent rate, because investment is not an easy task, and many factors need to be considered, including some factors that are difficult to predict with models.
\end{abstract}


\section{Introduction}
The certainty equivalent rate is a measure derived from the certainty equivalent\footnote{A detailed explanation of the certainty equivalent can be found in \cite{Investopedia(2021)}.}, which plays a pivotal role in financial investment. Investors usually need to refer to the changing trend of this value to decide whether a certain company is worth investing in or if certain companies are worth investing in, that is, it is used to determine the priority of investment. The main purpose of this paper is to solve these two problems.\par
First, we give the mathematical expression of the certainty equivalent rate using the put-call parity formula. From the mathematical expression, we can find the factors that cause changes in the certainty equivalent rate.\par
Second, we select the stocks and options of three companies with the best performance from the beginning of 2023 to the current time and the stocks and options of three companies with the worst performance for empirical research. Through data visualization, we obtain a general conclusion.\par
Third, we apply the general conclusions drawn to the US electric vehicle industry. Specifically, we select three well-known US electric vehicle companies and explore whether they are worth investing in from the perspective of the certainty equivalent rate and the priority of investment.\par
Finally, we summarize this paper and emphasize that investment is an extremely complicated matter that requires the consideration of many factors, not just the certainty equivalent rate. When many factors are considered, investors are more likely to make the optimal decision.

\section{Theoretical Support}
The key theorem that we will use is put-call parity. A detailed explanation of put-call parity can be found in \cite{Hull(2022)}. \cite{Hull(2022)} considers interest and dividends to be paid in accordance with continuous compounding, but in a real financial market, interest and dividends are more likely to be paid at specific points in time rather than every second. Thus, we prefer to use the discrete-compounding version of put-call parity when we consider problems in a real financial market. Hence, the following is the detailed mathematical expression of put-call parity that we will use in this paper:
\begin{align}\label{1}
C + \frac{K}{(1+r)^{T}} = P + \frac{S}{(1+q)^{T}},
\end{align}
where
\begin{align*}
\begin{cases}
T &\overset{\textrm{def}}{=}\textrm{The time to maturity},\\
C &\overset{\textrm{def}}{=}\textrm{A given company's call option price with respect to the maturity date},\\
P &\overset{\textrm{def}}{=}\textrm{A given company's put option price with respect to the maturity date},\\
K &\overset{\textrm{def}}{=}\textrm{A given company's option strike price with respect to the maturity date},\\
S & \overset{\textrm{def}}{=}\textrm{A given company's stock price with respect to the start date},\\
q & \overset{\textrm{def}}{=}\textrm{A given company's dividend yield},\\
r & \overset{\textrm{def}}{=}\textrm{A given company's put-call parity certainty equivalent rate}.
\end{cases}
\end{align*}
Based on (\ref{1}), we can obtain the mathematical expression for $r$:
\begin{align}
r = \left[\frac{K(1+q)^{T}}{S+(P-C)(1+q)^{T}}\right]^{\frac{1}{T}}.
\end{align}\label{2}
In the following sections, we will mainly use (\ref{2}) to explore the relationship between the time to maturity $T$, the moneyness $S/K$, and the implied company-specific put-call parity certainty equivalent rate $r$.

\section{Empirical Research}

\subsection{Preparation}
Before conducting our empirical research, we had to figure out how to obtain the values of the arguments in (\ref{2}):
\begin{itemize}
\item \textbf{Argument $S$}: Since our purpose is to explore the stocks' behavior in 2023, we choose the start date as January 3, 2023, which is the first business day in 2023. Therefore, $S$ in (\ref{2}) will be the company's stock price on January 3, 2023. We can find these values in every stock's ``historical data'' section on \href{https://finance.yahoo.com/}{Yahoo Finance}.

\item \textbf{Argument $q$}: We will consider the value of the ``forward annual dividend yield''; the relevant data can be found in the ``statistics'' section on \href{https://finance.yahoo.com/}{Yahoo Finance}.

\item \textbf{Arguments $T$ and $K$}: On \href{https://www.cboe.com/}{CBOE}, we can find a given company's stock option's strike $T$ and its maturity date. Then, we subtract the start date (January 3, 2023) from the maturity date and convert the result into years\footnote{This is because the time to maturity $T$ in the put-call parity expression has units of years, and we assume that one calendar year has 252 business days in this paper.}. Finally, we obtain the value of the time to maturity $T$. 

\item \textbf{Arguments $C$ and $P$}: Although we cannot obtain the values of these two arguments directly from \href{https://www.cboe.com/}{CBOE}, we can obtain the ``bid'' and ``ask'' of every option. Here, we calculate the mid-price of the bid and ask, and we consider it to be the corresponding call option price and put option price.
\end{itemize}

\subsection{Data Visualization and Explanation}
\begin{figure}[H]
  \centering
  \includegraphics[width=\textwidth]{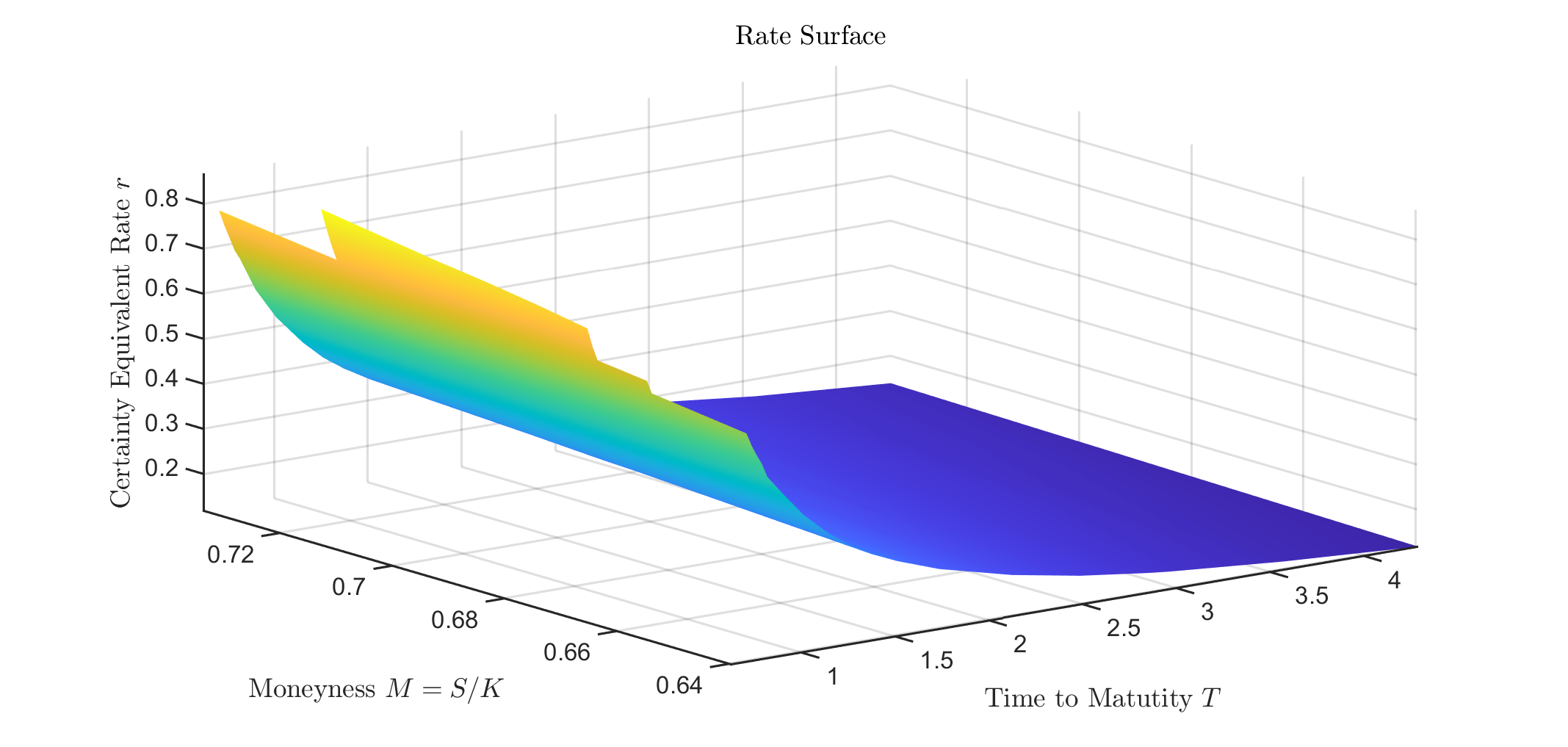}
  \caption{Relationship between the time to maturity, moneyness, and the implied put-call parity certainty equivalent rate for Apple stock}
  \label{AAPL}
\end{figure}

\begin{figure}[H]
  \centering
  \includegraphics[width=\textwidth]{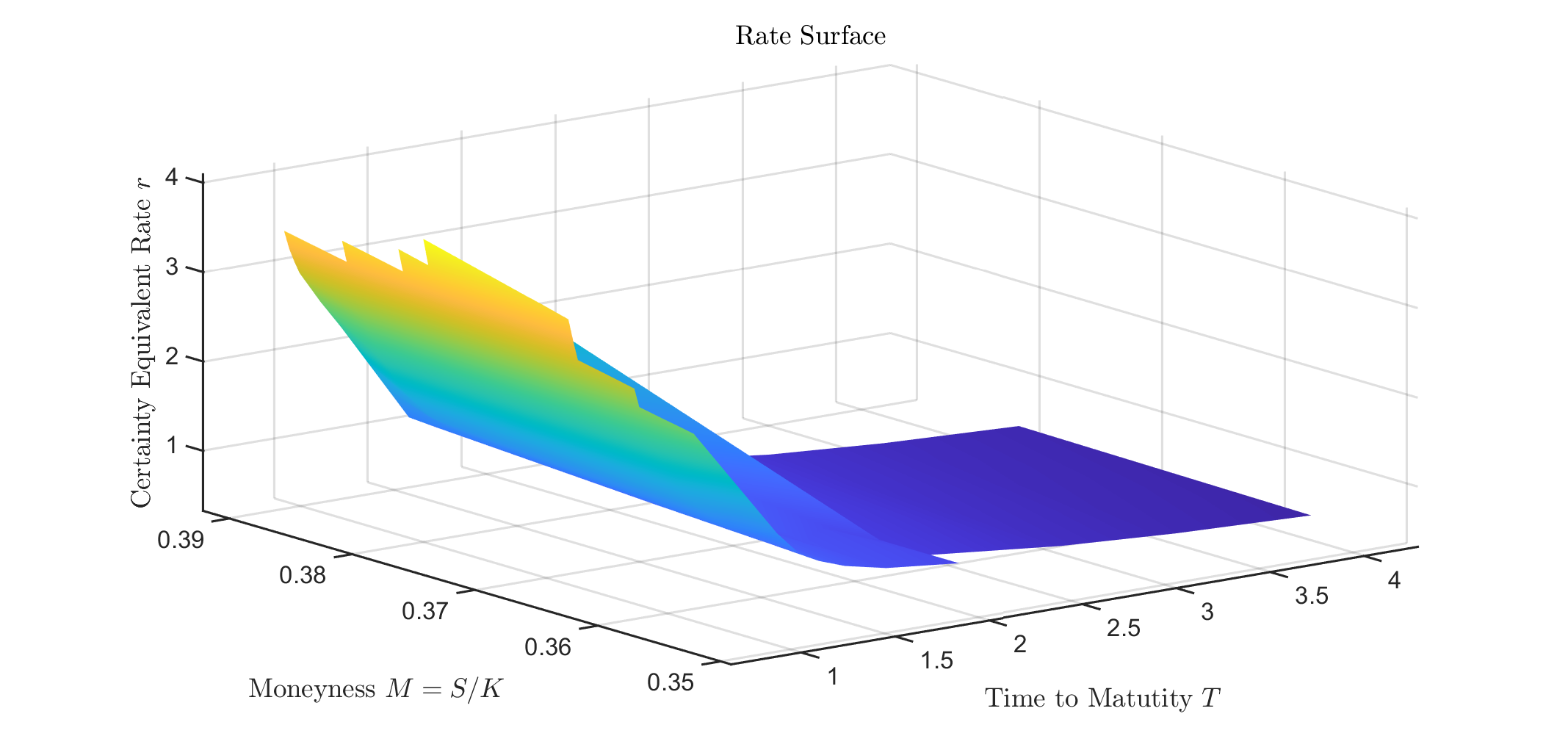}
  \caption{Relationship between the time to maturity, moneyness, and the implied put-call parity certainty equivalent rate for Nvidia stock}
  \label{NVDA}
\end{figure}

\begin{figure}[H]
  \centering
  \includegraphics[width=\textwidth]{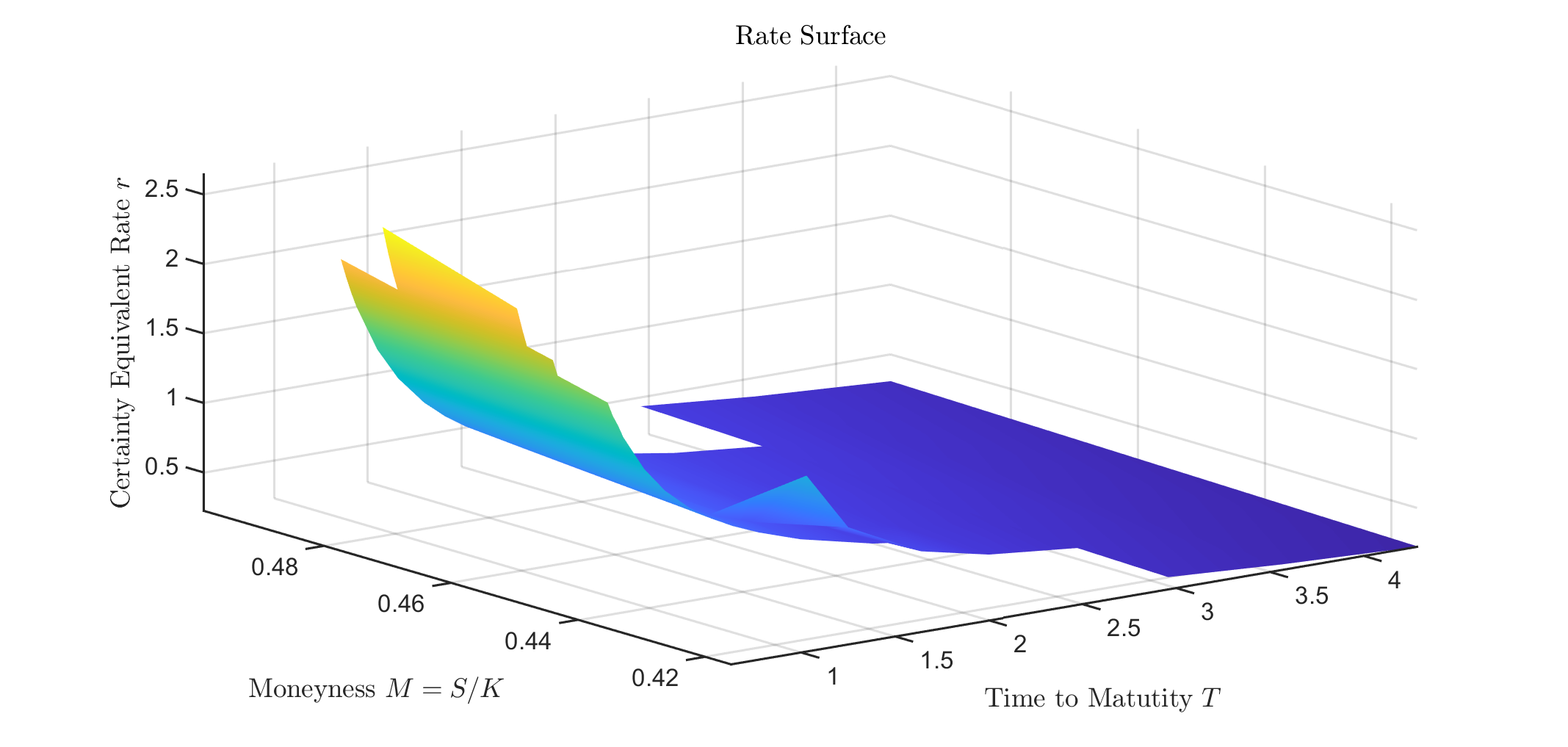}
  \caption{Relationship between the time to maturity, moneyness, and the implied put-call parity certainty equivalent rate for Meta stock}
  \label{META}
\end{figure}

\begin{figure}[H]
  \centering
  \includegraphics[width=\textwidth]{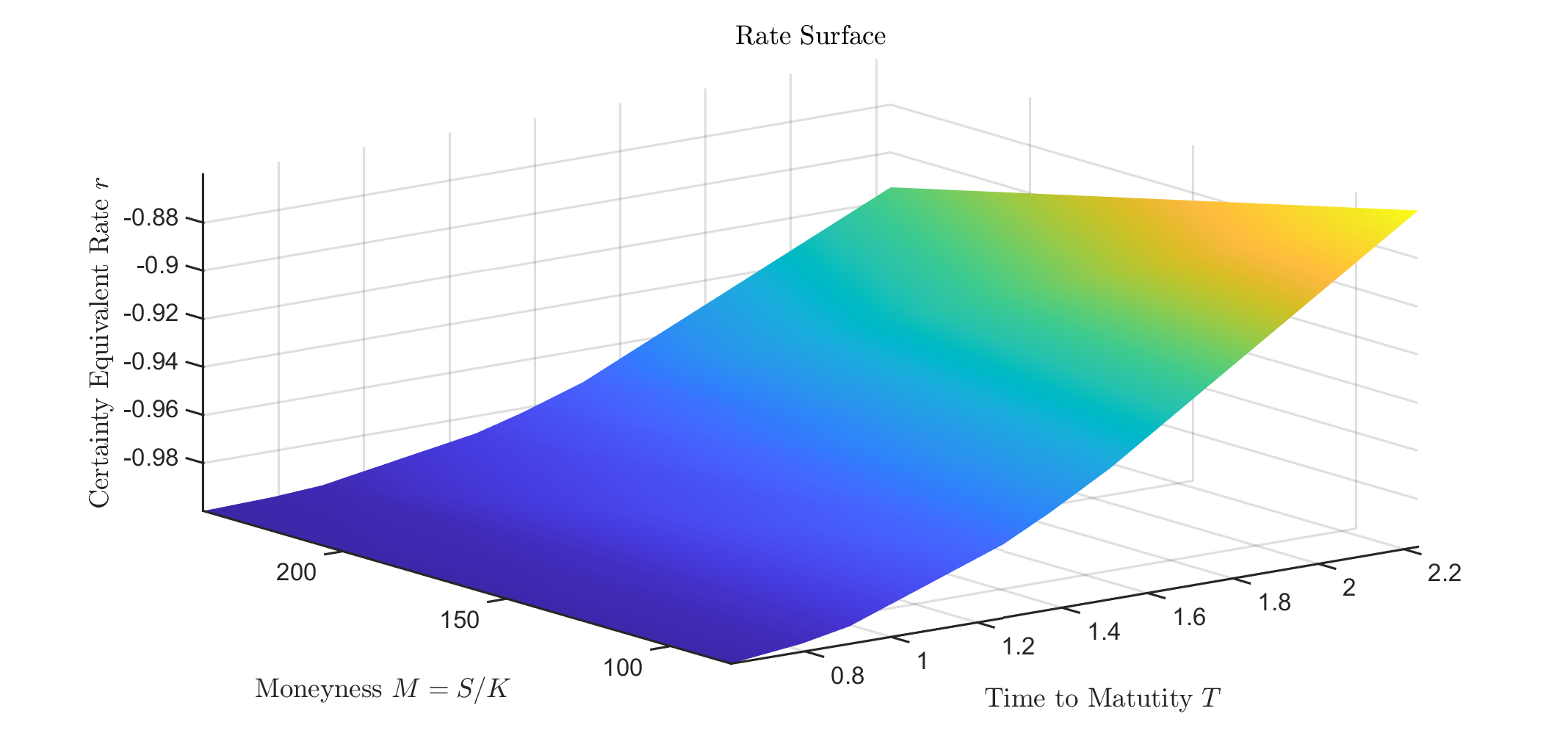}
  \caption{Relationship between the time to maturity, moneyness, and the implied put-call parity certainty equivalent rate for First Republic Bank stock}
  \label{FRCB}
\end{figure}

\begin{figure}[H]
  \centering
  \includegraphics[width=\textwidth]{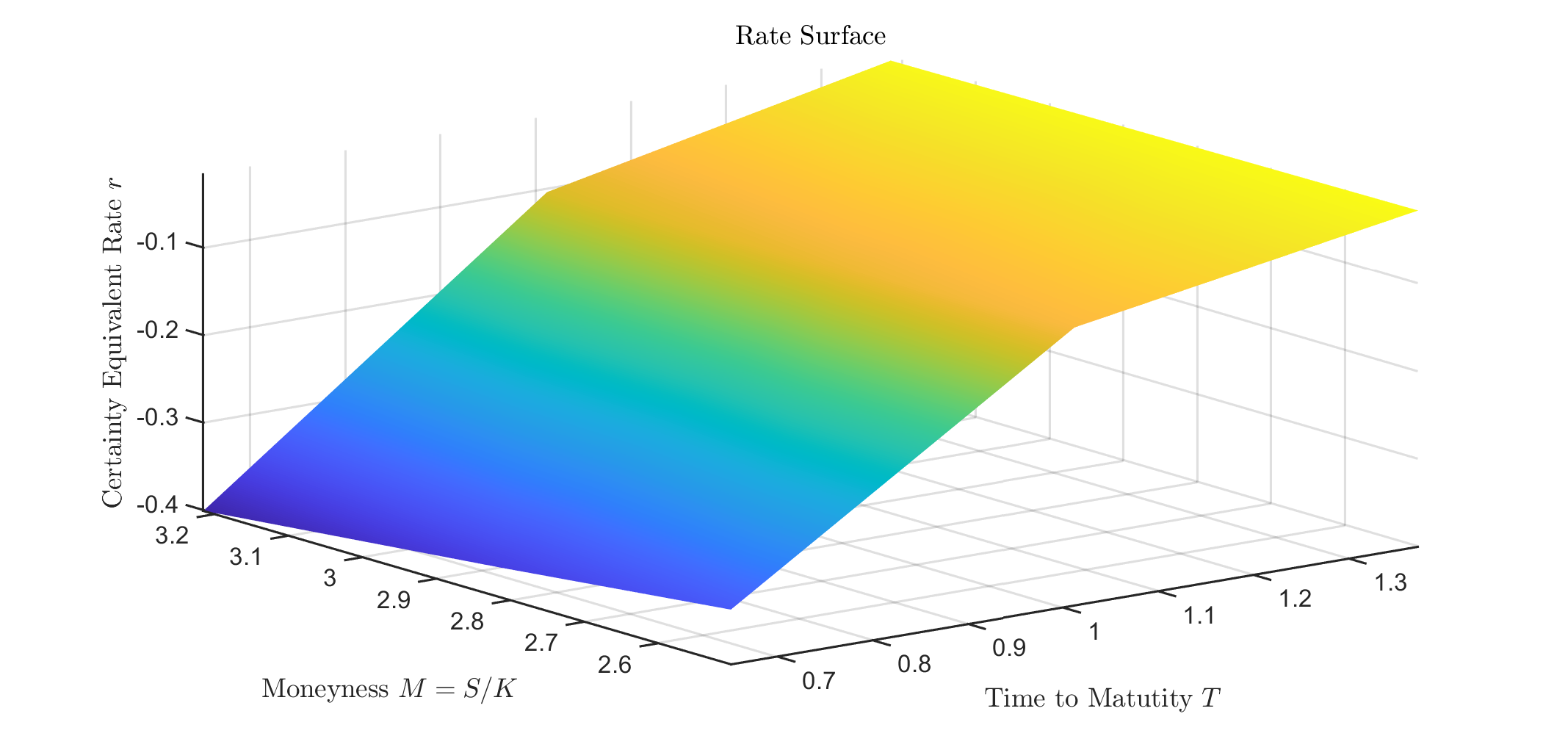}
  \caption{Relationship between the time to maturity, moneyness, and the implied put-call parity certainty equivalent rate for Signature Bank stock}
  \label{SBNY}
\end{figure}

\begin{figure}[H]
  \centering
  \includegraphics[width=\textwidth]{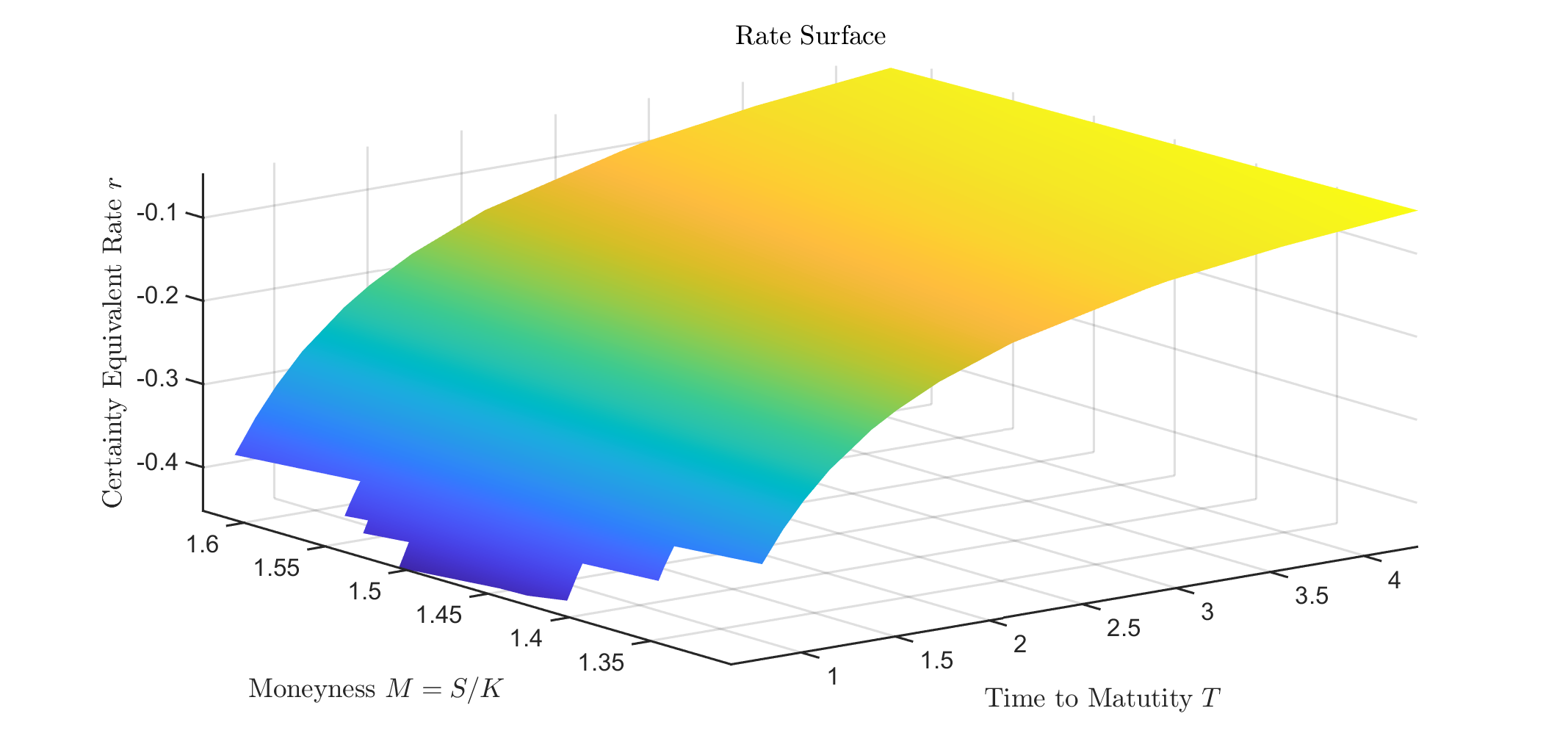}
  \caption{Relationship between the time to maturity, moneyness, and the implied put-call parity certainty equivalent rate for Charles Schwab stock}
  \label{SCHW}
\end{figure}

Based on financial news from \cite{Forbes(2023)}, \cite{CNBC(2023)}, \cite{Invesopedia(2023)}, and \cite{BENZINGA(2023)}, we can select three of the best-performing stocks, which come from \textbf{Apple}, \textbf{Nvidia}, and \textbf{Meta}, respectively. On the other hand, the three worst-performing stocks that we select are from \textbf{First Republic Bank}, \textbf{Signature Bank}, and \textbf{Charles Schwab}, respectively. Next, we use MATLAB to create figures that describe the relationship between the time to maturity $T$\footnote{The options' maturity dates for Apple, Nvidia, Meta, and Charles Schwab range from 06/09/2023 to 12/19/2025; the options' maturity dates for First Republic Bank range from 06/09/2023 to 07/19/2024; and the options' maturity dates for Signature Bank range from 06/16/2023 to 12/15/2023.}, the moneyness $S/K$\footnote{The strike prices for these six companies' options correspond to the maturity dates.}, and the implied put-call parity certainty equivalent rate $r$ for the companies we selected, and we explain some key values from these figures\footnote{Figures \ref{AAPL}--\ref{META} correspond to the best-performing companies, while Figures \ref{FRCB}--\ref{SCHW} correspond to the worst-performing companies.}.

\begin{table}[H]
\centering
\begin{tabular}{c c c c}
\hline
\textrm{\textbf{Company}} & \textrm{\textbf{Maximum value}} &\textrm{\textbf{Minimum value}} &\textrm{\textbf{Mean value}}\\ \hline
\textrm{Apple} & 0.8644 & 0.1181 & 0.4437\\
\textrm{Nvidia} & 4.0908 & 0.3308 & 1.8897\\
\textrm{Meta} & 2.6418 & 0.2238 & 1.2135 \\ \hline
\textrm{First Republic Bank} & -0.8600 & -0.9998 & -0.9731\\
\textrm{Signature Bank} & -0.0166 & -0.4008 & -0.1601\\ 
\textrm{Charles Schwab} & -0.0481 & -0.4523 & -0.2625\\ \hline
\end{tabular}
\caption{Implied put-call parity certainty equivalent rate (best- and worst-performing companies)}\label{Table 1}
\end{table}
From Table \ref{Table 1}\footnote{Via put-call parity, we can obtain different rates with respect to different time to maturity values, and thus we can determine the maximum value, minimum value, and mean value of these rates. The values in Table \ref{Table 2} are similar.}, it is clear that the best-performing companies have strictly positive implied put-call parity certainty equivalent rates, regardless of the maximum value, minimum value, or mean value. On the other hand, the worst-performing companies have strictly negative implied put-call parity certainty equivalent rates.\par
A positive rate means that investors have a high probability of obtaining a return if they invest in the company's stock, while a negative rate means that investors may lose their money if they try to invest in the company's stock. By combining the data and the financial news, we find that the implied put-call parity certainty equivalent rate is quite useful; it can help investors determine which company's stock option is worth investing in.\par
Now, let us consider the shape of the graph. For the best-performing companies, we can observe that when the time to maturity $T$ is small, the corresponding rate is high, which means that in the near future, the return of the stock option is quite high, so that investors may make money by investing in the option. Of course, they may have to take on some amount of risk. In this case, we would suggest that the investor consider Apple first, then Meta, and finally Nvidia. A high rate represents a high risk, and normal investors definitely do not want to take on a high risk when they decide to invest in something.\par
As time goes by, i.e., as the value of the time to maturity $T$ becomes larger, the implied put-call parity certainty equivalent rate will become smaller because it is reasonable to consider the long-term rate as the riskless rate of the financial market. Additionally, it is clear that the riskless rate of the financial market should be lower than the near-future implied put-call parity certainty equivalent rates of the best-performing companies.\par
Let us consider the worst-performing companies. We can see that the values of the certainty equivalent rates of these companies are negative, which means that investors have a high risk of losing money if they decide to invest in these companies' stock options.\par
Finally, from Figures \ref{AAPL}--\ref{SCHW}, we can see that if we fix the value of the time to maturity $T$ and change the value of the moneyness $S/K$, the value of the implied certainty equivalent rate $r$ hardly changes, which tells us that the implied certainty equivalent rate is almost independent of moneyness.

\section{Application: Electric Vehicle Industry}
\begin{figure}[H]
  \centering
  \includegraphics[width=\textwidth]{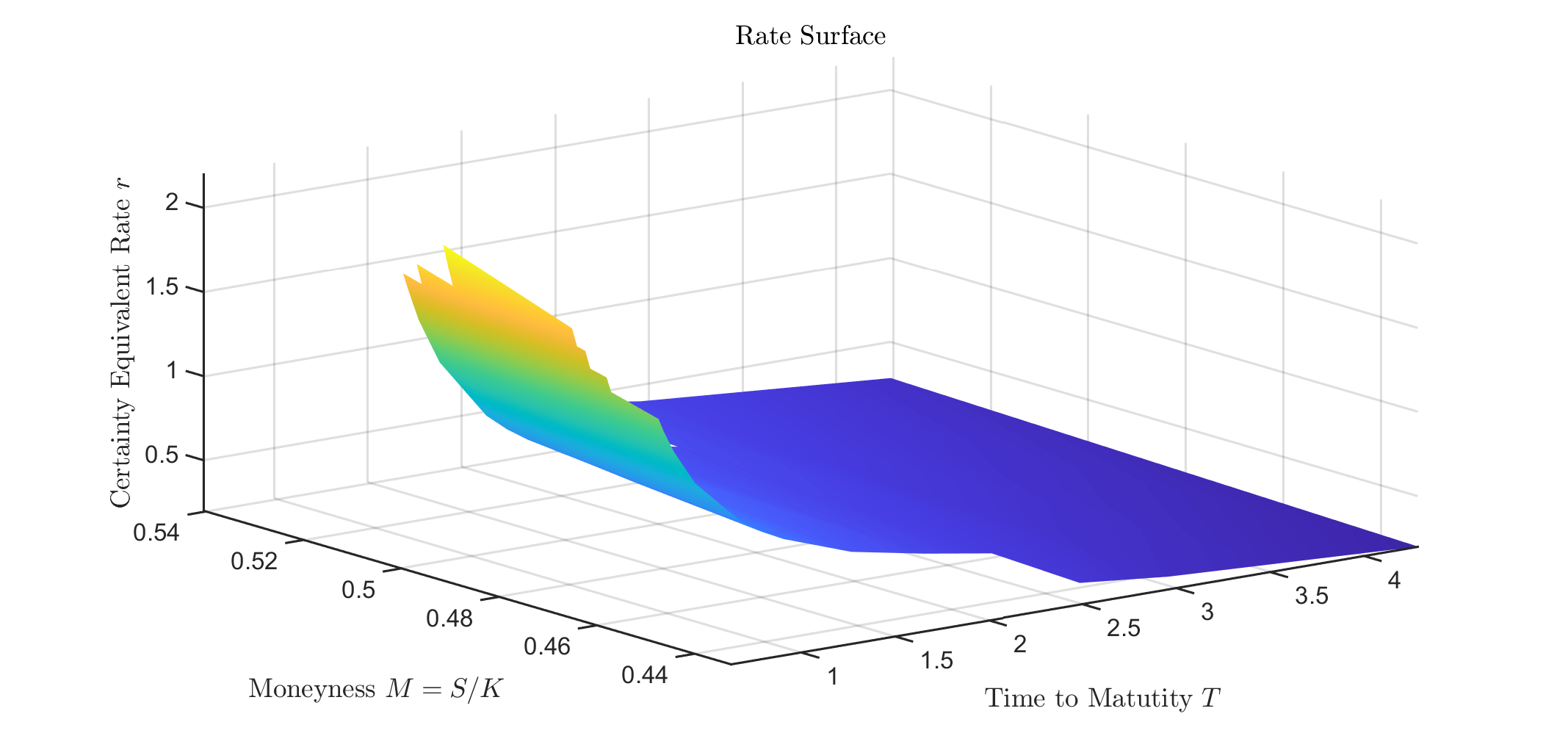}
  \caption{Relationship between the time to maturity, moneyness, and the implied put-call parity certainty equivalent rate for Tesla stock}
  \label{TSLA}
\end{figure}

\begin{figure}[H]
  \centering
  \includegraphics[width=\textwidth]{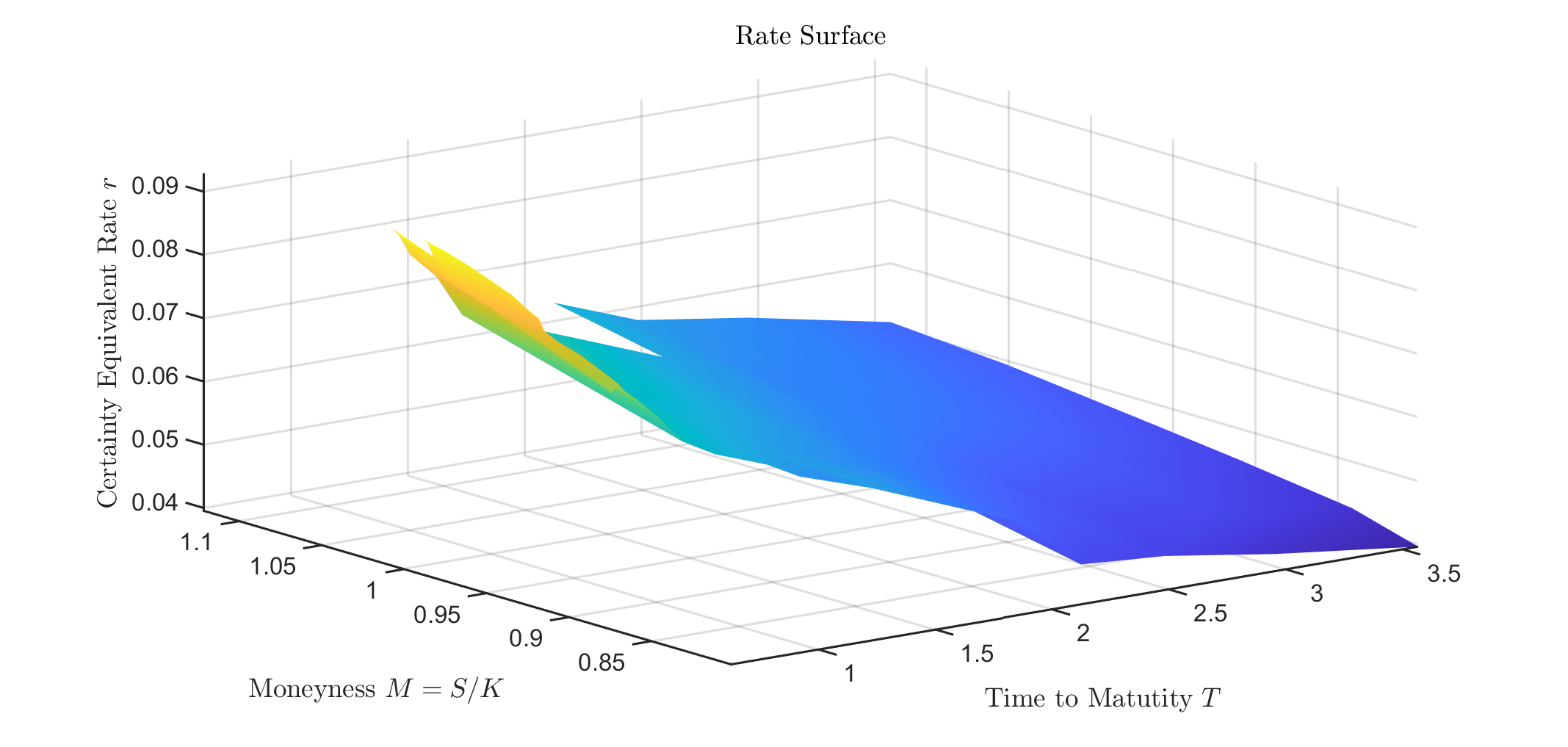}
  \caption{Relationship between the time to maturity, moneyness, and the implied put-call parity certainty equivalent rate for General Motors stock}
  \label{GM}
\end{figure}

\begin{figure}[H]
  \centering
  \includegraphics[width=\textwidth]{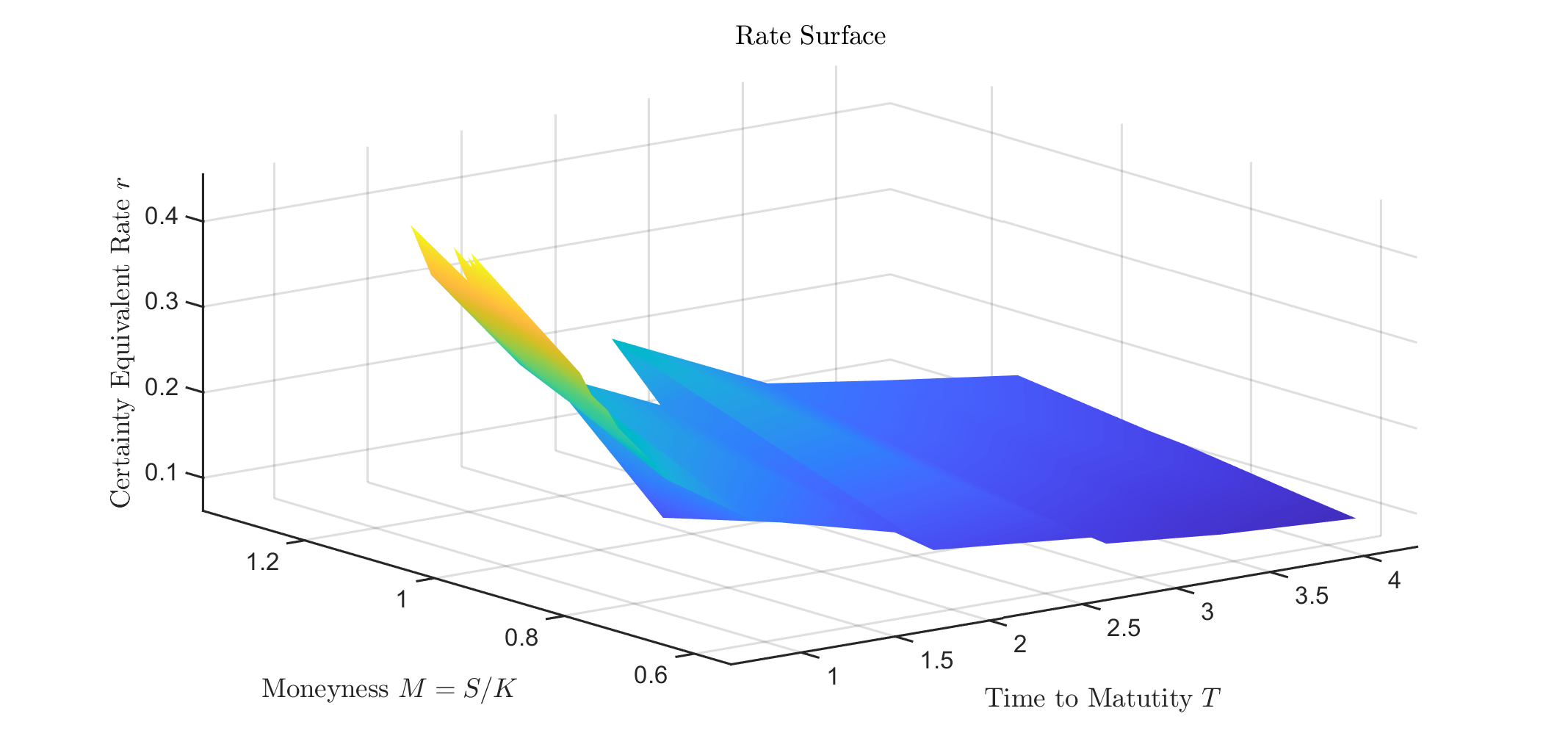}
  \caption{Relationship between the time to maturity, moneyness, and the implied put-call parity certainty equivalent rate for Ford Motor Company stock}
  \label{F}
\end{figure}

Today, more and more people are paying attention to environmental issues. To reduce the pollution released by vehicles, people are considering driving electric vehicles\footnote{The current electric vehicle market situation is described in \cite{Sanguesa et al.(2021)}, and the reasons that electric vehicles can reduce pollution can be found in \cite{Thomas(2012)}.} instead of traditional oil-powered vehicles. In the US, there are several well-known companies that produce electric vehicles, such as Tesla, General Motors, and Ford Motor Company\footnote{The options' maturity dates for Tesla and Ford Motor Company range from 06/09/2023 to 12/19/2025; the options' maturity dates for General Motors range from 06/09/2023 to 06/20/2025. The options' strike prices correspond to the maturity dates.}. We can apply the results we obtained in the previous section to these electric vehicle companies. Figures \ref{TSLA}--\ref{F} visualize the data of these companies, and the key values derived from these figures are given in Table \ref{Table 2}.

\begin{table}[H]
\centering
\begin{tabular}{c c c c}
\hline
\textrm{\textbf{Company}} & \textrm{\textbf{Maximum value}} &\textrm{\textbf{Minimum value}} &\textrm{\textbf{Mean value}}\\ \hline
\textrm{Tesla} & 2.1970 & 0.1994 & 1.0320\\
\textrm{General Motors} & 0.0927 & 0.0395 & 0.0708\\
\textrm{Ford Motor Company} & 0.4549 & 0.0615 & 0.2626 \\ \hline
\end{tabular}
\caption{Implied put-call parity certainty equivalent rate (electric vehicle companies)}\label{Table 2}
\end{table}
Based on the shapes of the graphs shown in Figures \ref{TSLA}, \ref{GM}, and \ref{F}, we can see that these three companies are all doing well. In the near future, we recommend that investors invest in these companies' stock options. According to the data from Table \ref{Table 2}, we can see that if we plot the three surfaces on the same coordinate axis, the order of these three surfaces from top to bottom is Tesla, then Ford Motor Company, and finally General Motors. Based on the positional relationships, we can see that investors should prefer to invest in General Motors, then Ford Motor Company, and finally Tesla.

\section{Summary}
In this paper, we have used put-call parity to derive the implied put-call parity certainty equivalent rate. We have also considered the meaning of positive rates and negative rates. Then, we utilize the idea\footnote{The implied volatility surface is a three-dimensional surface that explores the relationship between the time to maturity $T$, moneyness $S/K$, and volatility $\sigma$.} of the implied volatility surface to construct the implied put-call parity certainty equivalent rate surface. From the relative positions of these surfaces, we can determine which stock option we should consider investing in first.\par
However, the certainty equivalent rate is only one factor that should be considered in investing. It is obvious that investors cannot consider this factor alone. When deciding to invest in a product, investors should also consider other factors, such as political factors. To be more specific, at the end of May 2023, Tesla CEO Elon Musk's visit to China\footnote{See the details in \cite{Kharpal(2023)}.} caused Tesla's stock to soar and made him the world's richest man. Thus, investors who invested in Tesla before this event made a large amount of money. It is clear that we cannot predict such an outcome via a mathematical model. Hence, when deciding to invest in a product, investors should also consider other factors, so that they have a better chance of making optimal decisions.

\end{document}